\documentclass{aa}
\usepackage{epsfig}

\def\etal{{\rm et al.\thinspace}}
\def\eg{{\rm e.g.}}

\def\ie{{\rm i.e.\ }}
\def\cf{{\rm cf.\ }}

\def\spose#1{\hbox to 0pt{#1\hss}}
\def\ltsimm{\mathrel{\spose{\lower 3pt\hbox{$\sim$}}
	\raise 2.0pt\hbox{$<$}}}
\def\gtsimm{\mathrel{\spose{\lower 3pt\hbox{$\sim$}}
	\raise 2.0pt\hbox{$>$}}}
\def\Mdot{\hbox{${\dot M}$} \,}

\def\km{{\rm\thinspace km}}
\def\cm{{\rm\thinspace cm}}
\def\pc{{\rm\thinspace pc}}
\def\s{{\rm\thinspace s}}
\def\yr{{\rm\thinspace yr}}
\def\g{{\rm\thinspace g}}
\def\K{{\rm\thinspace K}}

\def\kmps{\hbox{${\rm\km\s^{-1}\,}$}}
\def\erg{{\rm\thinspace erg}}

\def\Msol{\hbox{${\rm\thinspace M_{\odot}}$}}
\def\Msolpyr{\hbox{${\rm\Msol\yr^{-1}\,}$}}
\def\ergpcm2ps{\hbox{${\rm\erg\cm^{-2}\s^{-1}\,}$}}
\def\pcm3{\hbox{${\rm\cm^{-3}\,}$}}
\def\gpcm3{\hbox{${\rm\g\cm^{-3}\,}$}}

\begin{document}
   
\title{The Formation of Broad Emission Line Regions in Supernova-QSO Wind
Interactions II. 2D Calculations}

\author{J.M. Pittard \inst{1}, J.E. Dyson \inst{1}, S.A.E.G. Falle \inst{2} and T.W. Hartquist \inst{1}
}

\institute{Department of Physics and Astronomy, The University of Leeds, 
        Woodhouse Lane, Leeds, LS2 9JT, UK
\and
Department of Applied Mathematics, The University of Leeds, 
        Woodhouse Lane, Leeds, LS2 9JT, UK
\\}

\offprints{J. M. Pittard, \email{jmp@ast.leeds.ac.uk}}

\date{Received date / Accepted date}

\abstract{One aspect of supernova remnant evolution that is 
relatively unstudied is the influence of an AGN environment.
A high density ambient medium and a nearby powerful continuum source
will assist the cooling of shocked ejecta and swept-up gas. Motion
of the surrounding medium relative to the remnant will also affect the
remnant morphology. In an extension to previous work 
we have performed 2D hydrodynamical calculations of 
SNR evolution in an AGN environment, and have determined the
evolutionary behaviour of cold gas in the remnant. The cold gas
will contribute to the observed broad line emission in AGNs, and 
we present preliminary theoretical line profiles from our calculations.
A more detailed comparison with observations will be performed in future
work. The SNR-AGN interaction may be useful as a diagnostic of the AGN 
wind.
}

\titlerunning{2D Calculations of Supernova-QSO Wind Interactions}
\authorrunning{Pittard, Dyson, Falle \& Hartquist}

\maketitle

\label{firstpage}

\section{Introduction}
\label{sec:intro}
Active galaxies produce 
prodigious luminosities in tiny volumes, display very strong cosmological 
evolution, and are known for the diversity of their behaviour 
(\eg, Osterbrock \& Matthews \cite{OM1986}). 
Their activity arises from the release of 
gravitational energy from accretion onto a supermassive 
($\sim 10^{6} - 10^{9} \Msol$) black hole and/or the cumulative effects 
of short-lived episodes of nuclear star formation, and is often accompanied
by nuclear winds and jets. There is evidence (\eg, Williams \& Perry 
\cite{WP1994}) that intense star formation and nuclear activity are 
related, and starbursts are one possible way to fuel the central black hole. 
In turn, the nuclear wind and the extreme radiation field act back on
the starburst components, influencing stellar winds (and thus
stellar evolution), wind blown bubbles, and supernova remnants.
For the most luminous active galactic nuclei (AGNs) to be powered by
accretion requires that the density of the interstellar medium
be enhanced, for example by galaxy encounters or mass loss from a 
dense nuclear stellar cluster.
One of the most important questions concerning the AGN phenomenon is 
therefore the connection between starburst and nuclear activity.

A defining characteristic of AGNs are their possession of strong (and often 
very broad) line emission. This provides detailed information on the
physical conditions right down into the AGN core, and over the years
a great deal has been learned about the properties of the gas comprising the 
broad emission line region (BELR). It is photoionized, since reverberation 
studies (\eg, Clavel \etal \cite{C1991}) show the direct response of emission 
line strengths to continuum variability. The absence of deep Ly$\alpha$ 
absorption indicates that the BELR covers only $5-25$\% of the continuum 
source (\eg, Bottorff \etal \cite{BKSB1997}). It has a small volume filling 
factor $\sim 10^{-7}$ (as determined from the observed line strength to 
continuum ratio; Netzer \cite{N1990}). It also generates a wide range 
of line profile shapes (indicating that the geometry and kinematics 
are complex and varied), and shows evidence of at least 
a two-component structure (Collin-Souffrin \etal
\cite{CDT1982}, \cite{CDJP1986}; Wills \etal \cite{WNW1985}). One of 
these components is associated with high ionization 
lines, including 
\ion{C}{iii}], \ion{C}{iv}, 
and other multiply 
ionized species, and is known as the HIL. The second component can be 
identified with the low ionization lines which include the bulk of the
Balmer lines, and lines of singly ionized species (\eg, \ion{Mg}{ii},
\ion{C}{ii}, \ion{Fe}{ii}) and is known as the LIL.

The regions emitting the LIL and HIL display different kinematics, as 
deduced from studies of the profiles and line widths 
(\eg, Gaskell \cite{G1988}; Sulentic \etal \cite{S1995}), and the HIL 
are systematically blue-shifted with respect to the LIL 
(see, \eg, Sulentic \etal \cite{SMD2000}). To 
account for the variability of the low ionization \ion{Mg}{ii} and
Balmer lines the LILs must be optically thick (\eg, 
Ferland \etal \cite{F1992}).  
On the other hand, optically thin gas may account for the Baldwin
effect (a negative correlation between the ultraviolet emission-line 
equivalent width and continuum luminosity), although it remains to be seen
if this is due to sample biases (Sulentic \etal \cite{SMD2000}), and 
for the Wamsteker-Colina effect (a negative correlation between between the 
\ion{C}{iv} $\lambda$1549/Ly$\alpha$ ratio and continuum luminosity;
Shields \etal \cite{SFP1995}).

It is now fairly clear that the Balmer lines form at the surface of an
accretion disk, or close to the surface in an accretion disk wind 
(\eg, Collin-Souffrin \etal \cite{CDMP1988}; Marziani \etal \cite{M1996};
Nicastro \cite{N2000}), and approximately three 
quarters of the total luminosity of the broad-line emission 
is estimated to arise in the LILs (Collin-Souffrin \etal \cite{CDMP1988}). 
The geometrical distribution and kinematics of the HIL gas is, however, 
much less clear: it may arise in a spherical outflow, but a 
biconical `jet-like' distribution is another possibility. BELR size
determinations are commonly obtained using reverberation and photoionization
techniques, though it was concluded in a recent study that
gravitational microlensing could be a useful alternative method, 
particularly for the HILs (Abajas \etal \cite{AMMPO2002}).

Many theoretical explanations have been proposed for the origin of
the BELR. They include: i) magnetic acceleration of clouds off accretion 
discs (Emmering \etal \cite{EBS1992}); ii) the interaction 
of an outflowing wind with the surface of an accretion 
disc (Cassidy \& Raine \cite{CR1996}); iii) interaction of stars with 
accretion discs (Zurek \etal \cite{ZSC1994}); iv) 
tidal disruption of stars in the gravitational field of the BH 
(Roos \cite{R1992}); v) interaction of an AGN wind with supernovae and 
star clusters (Perry \& Dyson \cite{PD1985}; Williams \& Perry \cite{WP1994}); 
vi) emission from accretion shocks (Fromerth \& Melia \cite{FM2001});
vii) ionized stellar envelopes 
(\eg Torricelli-Ciamponi \& Pietrini \cite{TP2002}). 
Many other models have been shown to possess serious difficulties
(see references in Pittard \etal \cite{PDFH2001}): in particular, any model 
must overcome the `confinement problem', and/or continually generate clouds.

Many of the mechanisms on which the various models are based will probably
contribute to the production of the
observed BELR gas. However, it is clear that some of the proposed mechanisms
will be more dominant than others, at least under certain conditions. For 
example, the rate of tidally disrupted stars in high luminosity AGNs is
likely to be too low to account for much of the BELR in these objects. 
Which are generally the dominant contributions remains, to date, largely 
unknown.

Even if the SNR-AGN wind interaction is not the dominant formation
mechanism of the HIL gas, it is of interest due to its potential as a 
diagnostic of the AGN wind. Supernovae must occur close to the central
AGN engine - in our own Galaxy there exists a cluster of a few dozen
evolved massive stars with initial masses $M > 20 \Msol$ (see 
Figer \& Kim \cite{FK2002} and references therein) in a region of
1.6~pc diameter centered on Sgr~A*. Recent absorption line
measurements from one of the high velocity stars are consistent 
with an O8-B0 dwarf with a mass $\sim 15 \Msol$ and a highly 
eccentric orbit which brings it within 1900~AU 
($\approx 6.3 \times 10^{-4} \pc$) of the supermassive black hole
(Ghez \etal \cite{G2003}). We can expect a similar, if not more
extreme, situation in the central regions of AGN.
 
The evolution of SNRs in a high density static ambient medium has
been studied by Terlevich \etal (\cite{TTFM1992}), 
with particular application to the formation of BELRs in starburst models 
developed to obviate the existence of supermassive black holes in 
the centres of AGNs. More recently, the additional influence of an intense 
continuum radiation field on the evolution of SNRs has been 
examined (Pittard \etal \cite{PDFH2001}). With Compton cooling and 
heating processes included in these calculations, 
the powerful flux of ionizing radiation influences
the thermal evolution of shocked regions. A central finding 
was that shocked gas could radiate efficiently enough to 
cool to temperatures and densities appropriate for the HIL. 

In this paper we present calculations which 
extend the work of Pittard \etal (\cite{PDFH2001}).
We describe the results of 2D axisymmetric hydrodynamical models of the 
interaction of an AGN wind with a supernova remnant. As the
formation, evolution, and structure of cold gas is of particular
interest, we have determined the mass of cool gas as a function of
time and present some simple modelling of line profiles. In a future
paper, we will perform more detailed line profile modelling and will
compare the results closely to observations.

In Sec.~\ref{sec:details} we discuss the details of our calculations; in 
Sec.~\ref{sec:results} we discuss our results; in 
Sec.~\ref{sec:line_profiles} we present some preliminary line profile
modelling; and in Sec.~\ref{sec:conclusions} we summarize and 
discuss future work.


\section{Details of the calculations}
\label{sec:details}
We performed our calculations with an adaptive grid hydrodynamical 
code which is second order accurate in space and time 
(see, \eg, Falle \& Komissarov \cite{FK1996}, \cite{FK1998}). 
We used a Chevalier-Nadyozhin similarity solution specified by 
$n=12$, $\delta=0$\footnote{$\delta=0$
specifies expansion into a constant density ambient medium. 
In reality the ejecta initially expands into circumstellar gas
expelled from the progenitor by its wind and by any outbursts.
The high ambient pressures considered in this work will confine
this material to a smaller volume than is the case for massive 
stars in a typical Galactic environment. For instance, with canonical values of
$\Mdot = 10^{-6} \Msolpyr$, and $v = 2000 \kmps$ for the 
progenitor wind, and $n = 10^{6} \pcm3$ and $T = 10^{7} \K$ for the
surrounding environment, the ram pressure of the progenitor wind is balanced
by the thermal pressure of the surrounding environment  
at a distance of $8.5 \times 10^{14} \cm$ ($2.75 \times 10^{-4} \pc$) 
from the center of the 
progenitor. If the ambient medium flows past the progenitor, 
substantially smaller conefinement radii can occur on the
upstream side. For the same parameters as above and $v = 3000 \kmps$ 
for the AGN wind, ram pressure balance occurs at a distance of 
$10^{14} \cm$ ($3.2 \times 10^{-5} \pc$) 
from the center of the progenitor. These 
distances are much smaller than the size of the subsequent
remnants (see Sec.~\ref{sec:results}), thereby justifying our 
assumption of a constant density surrounding medium and $\delta = 0$.
We note, however, that a more thorough treatment would consider the effect
of the mass of the confined wind on the results presented in
Sec.~\ref{sec:results}, since this could be significant compared 
to the amount of mass in the surrounding medium which is swept 
up by the time that the ejecta core interacts with the reverse shock.}
  as the initial profile for the SNR 
(Chevalier \cite{C1982}; Nadyozhin \cite{N1985}). 
In the majority of our calculations we assumed a canonical 
explosion energy of $10^{51}$~ergs and ejecta 
mass of $10 \Msol$, which is typical of a type~II SN. 
75 per cent of the mass and 58 per cent of the explosion energy are 
contained within a constant density core.
Heating and cooling rates for a canonical AGN spectrum were kindly supplied 
by Tod Woods (\cf Woods \etal \cite{W1996}) and are included in our
calculations. They are valid in the optically thin, low-density 
regime with solar abundances.
Further details and assumptions can be found in 
Pittard \etal (\cite{PDFH2001}) and references therein.

The thermal equilibrium of gas irradiated by the intense continuum
in an AGN may be described in terms of several ionization parameters.
When cold and hot phases exist with comparable pressure, it is convenient
to use a definition based on the ratio of the ionizing photon pressure
to gas pressure (\cf Krolik, McKee \& Tarter \cite{KMT1981}):

\begin{equation}
\label{eq:chi_ip}
\Xi = \frac{1}{4 \pi r^{2} n c k T} \int_{\nu_{\rm L}}^{\infty} L_{\nu} d\nu,
\end{equation}

\noindent where $r$ is the distance to the central continuum source, $n$
is the gas density, $T$ the gas temperature, $L_{\nu}$ the differential
luminosity, $\nu_{\rm L}$ the frequency at the Lyman limit, and the other
symbols have their usual meanings. Another common definition is the
ratio of the ionizing photon density to the gas density,

\begin{equation}
\label{eq:u_ip}
U = \frac{1}{4 \pi r^{2} n c} \int_{\nu_{\rm L}}^{\infty} 
\frac{L_{\nu}}{h\nu} d\nu.
\end{equation}

\noindent The relation between these two ionization parameters depends
on the shape of the ionizing spectrum. For the canonical AGN spectrum
in Cloudy\footnote{Ferland \cite{F2001}} 
(see Woods \etal \cite{W1996}), which we adopt for this work,

\begin{equation}
\label{eq:u_chi}
U = 0.0182 T_{4} \Xi,
\end{equation}

\noindent where $T_{4}$ is the gas temperature in units of $10^{4}$K.
Comparison with Roos (\cite{R1992}) shows that our adopted spectrum is
neither particularly soft or hard.

\begin{figure}[t]
\begin{center}
\psfig{figure=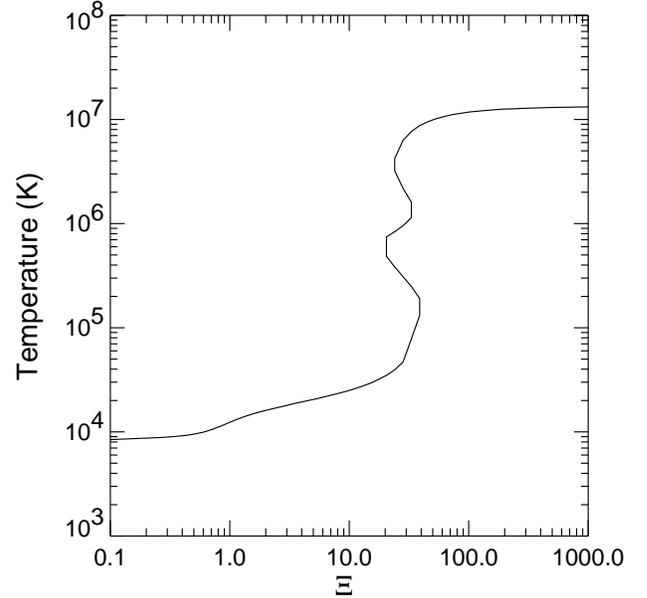,width=8.5cm}
\end{center}
\caption[]{Thermal equilibrium curve for the standard AGN spectrum in 
Cloudy (see Woods \etal \cite{W1996}).}
\label{fig:teq}
\end{figure}

In Figure~\ref{fig:teq} we show the thermal equilibrium curve for the 
assumed AGN spectrum as a function of temperature and ionization 
parameters, $\Xi$ and $U$. At low temperatures photoionization heating and 
cooling due to line excitation and recombination are in near balance.
At high temperatures, the equilibrium arises from the balance of Compton
heating and cooling. The exact shape of the thermal equilibrium 
curve at intermediate temperatures is a complicated function of the 
irradiating spectrum, the assumed abundances and thermal processes 
(\cf Krolik \etal \cite{KMT1981}), and varies substantially 
from source to source. Since we are not modelling a specific object,
we do not concern ourselves with the details of this part of the
equilibrium curve.

To obtain cool gas in thermal equilibrium we require ionization
parameters $\Xi \ltsimm 10$. 
As noted by Perry \& Dyson (\cite{PD1985}), shocked gas cooled back to
equilibrium can have a value of $\Xi$ much lower than its
pre-shock value. This is because the post-shock density and 
pressure can be much greater than the
pre-shock value. Therefore, strong shocks can create conditions
for the gas to cool to temperatures much lower than the surrounding
ambient temperature. 
The crucial question is whether the shocked gas remains at high
densities and pressures for long enough to cool from its post-shock
temperature to $T \sim 2 \times 10^{4}$~K. In our earlier 1D work 
(Pittard \etal \cite{PDFH2001}) we demonstrated that this was possible
for a supernova in a characteristic AGN environment.

In the next section we present results from 2D simulations of a 
SNR evolving in an AGN environment. We have computed models with 
ambient densities $n = 10^{5},10^{6} \pcm3$ and with
AGN wind speeds $v = 0,3000,5000,7000 \kmps$. The initial radius, 
expansion speed, and age of the SNR in our models is specified in 
Table~\ref{tab:init_radius_age} for each of the ambient densities. 
All models have the same ionization parameter and temperature for the 
ambient medium ($\Xi \approx 150$, $T = 1.33 \times 10^{7}$~K) unless
otherwise stated. We further assumed that the
central continuum source is distant enough that the flux of
ionizing radiation is constant over our computational volume. The 
SNR was evolved until the pressure of the shocked gas drops to the
point where it is no longer able to exist in the
cool phase ($T \sim 10^{4}$~K). As the remnant expands we periodically
regrid our model to a coarser set of grids.

\begin{table}
\begin{center}
\caption{Initial radius, $R$, expansion speed, $v_{\rm exp}$, and age, $t$, 
of the SNR for models with $E = 10^{51}$~erg and $M = 10 \Msol$, 
as a function of the ambient density, $n$.}
\label{tab:init_radius_age}
\begin{tabular}{lllll}
\hline
\hline
$n (\pcm3)$ & $R (10^{-3}\;{\rm pc})$ & $v_{\rm exp} (\kmps)$ & $t$~(yr) \\
\hline
$10^{5}$ & 3.4 & 11,000 & 0.3 \\
$10^{6}$ & 1.2 & 12,000 & 0.1 \\
\hline
\end{tabular}
\end{center}
\end{table}

\section{Results}
\label{sec:results}
 
\subsection{Expansion into a stationary medium}
\label{sec:stationary}
\begin{figure*}[h]
\begin{center}
\psfig{figure=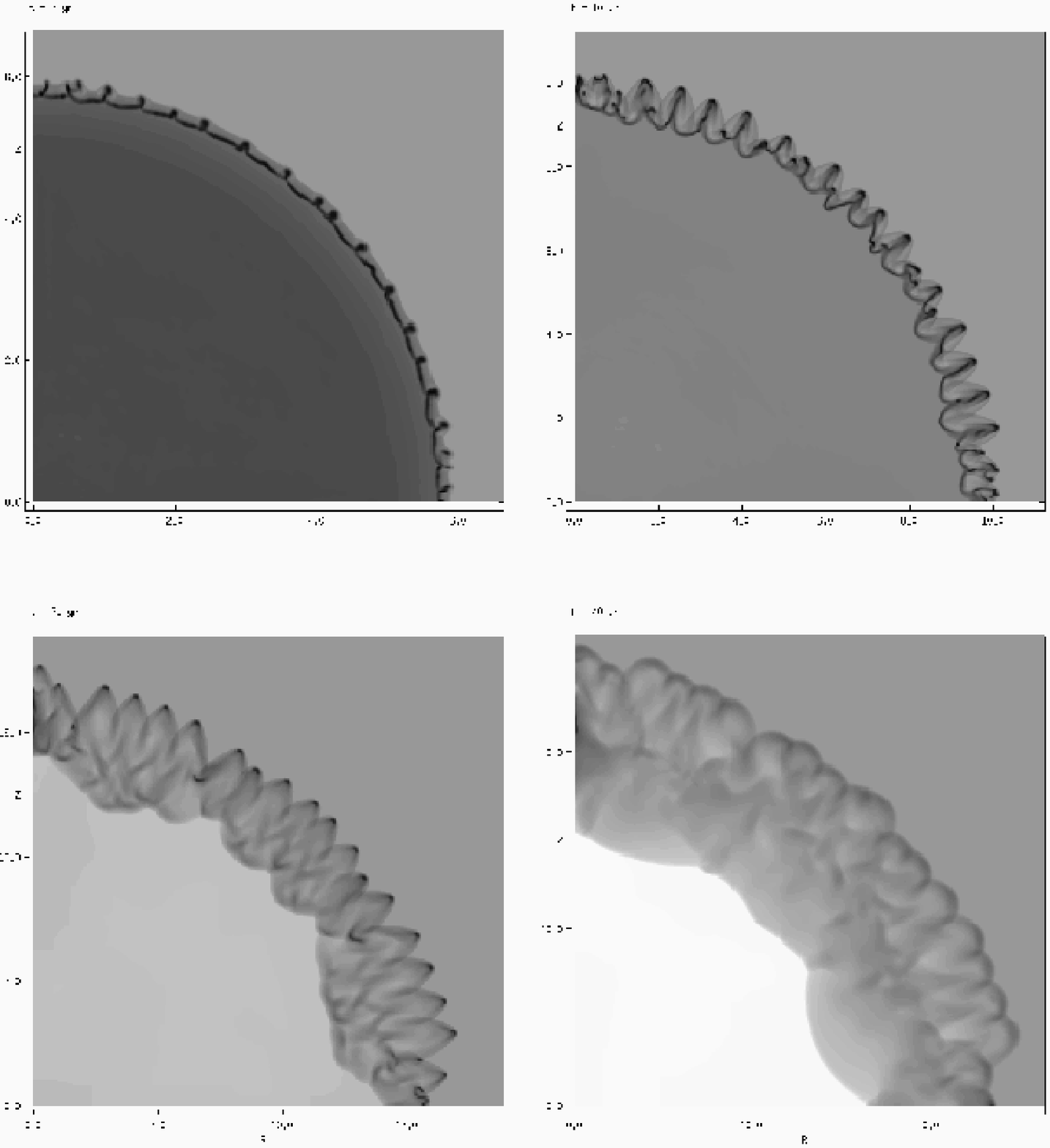,width=17.0cm}
\vspace{-15mm}
\end{center}
\caption[]{The evolution of a SNR expanding into a stationary environment 
of density $n = 10^{6} \pcm3$. Shown are logarithmic plots of the density with
a scale from $2.4 \times 10^{4} \pcm3$ (white) to $3.1 \times 10^{8} \pcm3$
(black). The age of the remnant is shown in the upper left corner of each 
plot. Note the change in the axis scaling between the plots. The unit of
length in these and subsequent plots is $1.1 \times 10^{16} \cm$ 
(0.0036~pc), and the ionization parameter of the surrounding medium,
$\Xi \approx 150$, unless explicitly stated otherwise.}
\label{fig:nw6_rho}
\end{figure*}

\begin{figure*}[h]
\begin{center}
\psfig{figure=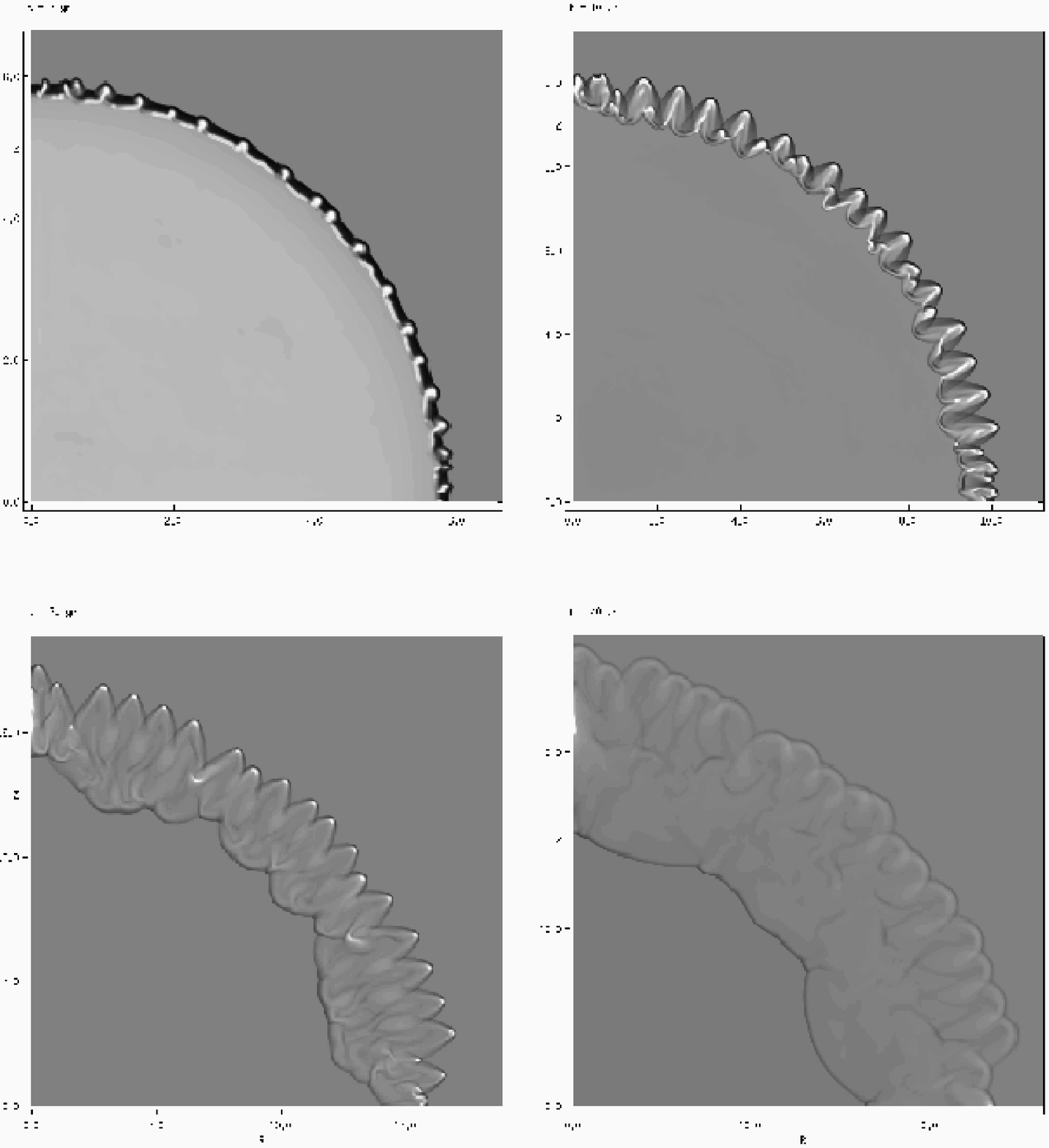,width=17.0cm}
\vspace{-15mm}
\end{center}
\caption[]{As Fig.~\ref{fig:nw6_rho} but showing logarithmic plots of the 
temperature with a scale from $10^{6} \K$ (white) to $1.58 \times 10^{8} \K$
(black). Note the creation, evolution, and destruction of cool gas 
($T < 10^{6} \K$) within the remnant.}
\label{fig:nw6_temp}
\end{figure*}

In Figs.~\ref{fig:nw6_rho} and~\ref{fig:nw6_temp} we show the evolution 
of a SNR expanding into a stationary environment with $n = 10^{6} \pcm3$. 
The shocked gas rapidly loses energy, first through inverse Compton 
scattering, and then through free-free and line cooling,
and is compressed into a relatively thin zone with the unshocked
ejecta dominating the remnant volume. The shocked gas has a fractional 
thickness of 0.132 in the adiabatic self-similar solution
(Chevalier \cite{C1982}; Nadyozhin \cite{N1985}), but at $t = 5 \yr$ the 
radiative energy loss has reduced this to just 0.036.

Since the shocked ejecta are denser than the swept-up gas, they cool quicker and 
form a thin dense shell within the narrow region of shocked gas, which is
bounded on its interior surface by the reverse shock. Since the dense
shell is decelerated by the hot gas on its outside surface, it
is subject to the Rayleigh-Taylor instability, and there have been many 
investigations of this behaviour in SNRs (\eg, Chevalier \etal \cite{CBE1992};
Chevalier \& Blondin \cite{CB1995}; Jun \etal \cite{JJN1996}; 
Blondin \etal \cite{BBR2001}; Blondin \& Ellison \cite{BE2001}; 
Wang \& Chevalier \cite{WC2002}).

While previous numerical calculations of remnants in the literature showed that the
shocked ejecta were unable to distort the position of the forward shock (the
Rayleigh-Taylor (R-T) ``fingers'' being limited to about half the thickness
of the region of hot, swept-up gas; Chevalier \etal (\cite{CBE1992}), 
a number of exceptions have recently been found. 
For instance, the existence of circumstellar cloudlets was found to 
enhance the growth of the R-T fingers by generating vortices 
in the swept-up gas (Jun \etal \cite{JJN1996}). 
Alternatively, if the forward shock is an efficient site for particle 
acceleration, the shock compression ratio is increased and the region
of hot, swept-up gas is reduced in thickness, allowing the convective
instabilities to reach all the way to the forward shock (Blondin
\& Ellison \cite{BE2001}). Finally, if the ejecta are clumpy, 
the greater momentum of the clumps enables them to push through the 
position of equlibrium pressure balance and to perturb (or puncture) the 
forward shock (\cf Blondin \etal \cite{BBR2001};
Wang \& Chevalier \cite{WC2002}). 

To this list we can add the present work.
In the high density, high radiation flux environment which we consider,
the region of swept-up gas is significantly more efficient at radiating
energy than in less extreme environments, and is thus more strongly compressed
than normal. In this sense our models mimic the
higher compression ratios found in models with efficient particle acceleration
(Blondin \& Ellison \cite{BE2001}), and like them we find that the 
high density ``fingers'' are able to distort the forward shock. 
Chevalier \etal (\cite{CBE1992}) had previously suggested the possibility
of a highly radiative inner shock front to explain protrusions seen in
VLBI observations of SN~1986J. The large decrease in entropy of the shocked 
ejecta greatly enhances this instability.

At $t = 10 \yr$, the R-T ``fingers'' have grown so long, and distorted the
forward shock to such an extent, that the instability begins to resemble
the non-linear thin shell instability (hereafter NTSI; Vishniac \cite{V1994}). 
To our knowledge this has never been seen before in SNRs, although it is a 
common phenomena in simulations of wind blown bubbles
(\eg, Garc\'{i}a-Segura \etal \cite{GML1996}) and colliding stellar winds
(\eg, Stevens \etal \cite{SBP1992}; Pittard \etal \cite{PSCI1998}).

We note that the ``fingers'' do not form at constant intervals along the
thin shell. Since they are not deliberately seeded, they naturally
develop from noise within the code itself. It is known from both analytical and
numerical work that 
small wavelength modes grow most rapidly for both the R-T instability 
(Chandrasekhar \cite{C1961}; Youngs \cite{Y1984}) and the NTSI
(Vishniac \cite{V1994}; Blondin \& Marks \cite{BM1996}).
However, it is extremely difficult to
resolve modes of the thin-shell instability, as a sufficient number of grid
cells across the thin shell is needed to follow the tangential flow of
material (Mac Low \& Norman \cite{MN1993}). Given that this is not the case 
in our models, we expect the development of this instability to differ in 
higher resolution runs. Nevertheless, we do not expect it to drastically alter the
mass of gas that cools (see Section~\ref{sec:cool_gas}).

Until about $t = 7 \yr$, the ejecta envelope impacts the reverse shock, 
and the pre-shock density remains relatively constant at 
$n \approx 7 \times 10^{6} \pcm3$.
However, soon after this the reverse shock interacts with the ejecta core.
During this stage the ram pressure of the ejecta on the shocked gas rapidly 
declines as the pre-shock density now decreases as $t^{-3}$, and the shocked region 
depressurizes as the reverse shock travels towards the center of the 
remnant. This loss in pressure raises the equilibrium ionization 
parameter, $\Xi$, of the shocked gas to the 
point that the equilibrium temperature changes from $\sim 10^{4} \K$ to 
$\sim 10^{7} \K$. Gas that has managed to cool to $T \sim 10^{4} \K$ is then 
heated. This behaviour is seen in Fig.~\ref{fig:nw6_temp}. At $t = 5 \yr$ 
much of the shocked ejecta has cooled to $T \approx 2 \times 10^{4} \K$, 
and at $t = 10 \yr$, this has significant structure from the action 
of the instabilities. However, by
$t = 20 \yr$ most of the cool gas has disappeared, with only a few regions 
concentrated at the densest points of the remnant, these typically 
being at the extremeties resulting from the ``fingers'' mentioned earlier. 
By $t = 40 \yr$ even these regions have disappeared, and the remnant 
is at an almost uniform temperature $T \approx 1.3 \times 10^{7} \K$.

\begin{figure*}[t]
\begin{center}
\psfig{figure=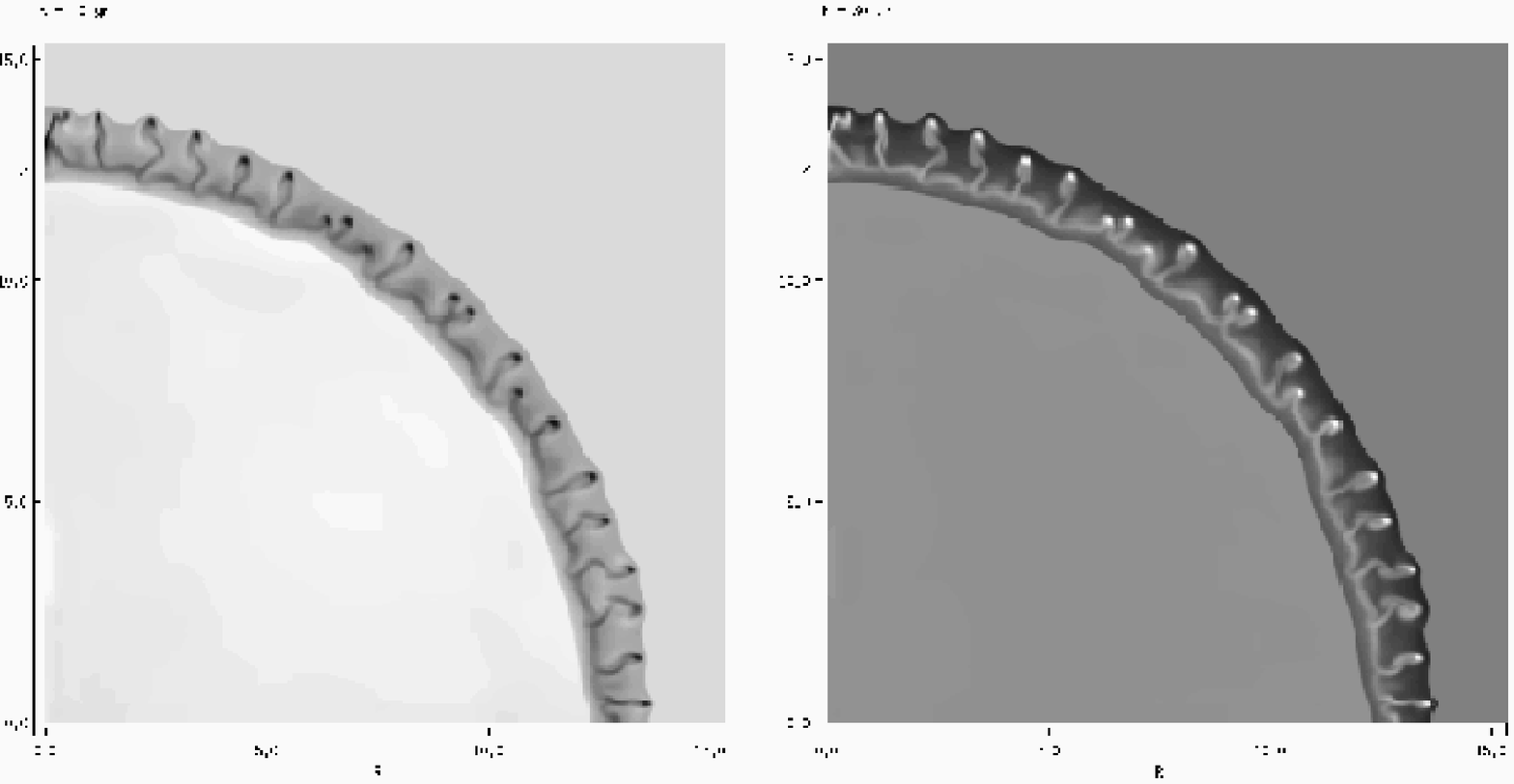,width=17.0cm}
\vspace{-15mm}
\end{center}
\caption[]{The morphology of a SNR expanding into stationary surroundings with 
$n = 10^{5} \pcm3$ at $t = 30 \yr$. Left: logarithmic density (grayscale from 
$4.7 \times 10^{4} \pcm3$ (white) to $1.8 \times 10^{7} \pcm3$ (black)). 
Right: logarithmic temperature (grayscale from $10^{6} \K$ (white) to 
$1.58 \times 10^{8} \K$ (black)). The unit of length is
$2.24 \times 10^{16} \cm$ (0.0073~pc).}
\label{fig:nw5}
\end{figure*}

Remnants expanding into lower density surroundings are characterized by a 
thicker region of swept up material; those expanding into higher density 
have a thinner region of swept up material. Fig.~\ref{fig:nw5} shows
the morphology of a remnant expanding into a stationary medium with 
$n = 10^{5} \pcm3$.

\subsection{Expansion into an AGN wind}
\label{sec:wind}
\begin{figure*}[t]
\begin{center}
\psfig{figure=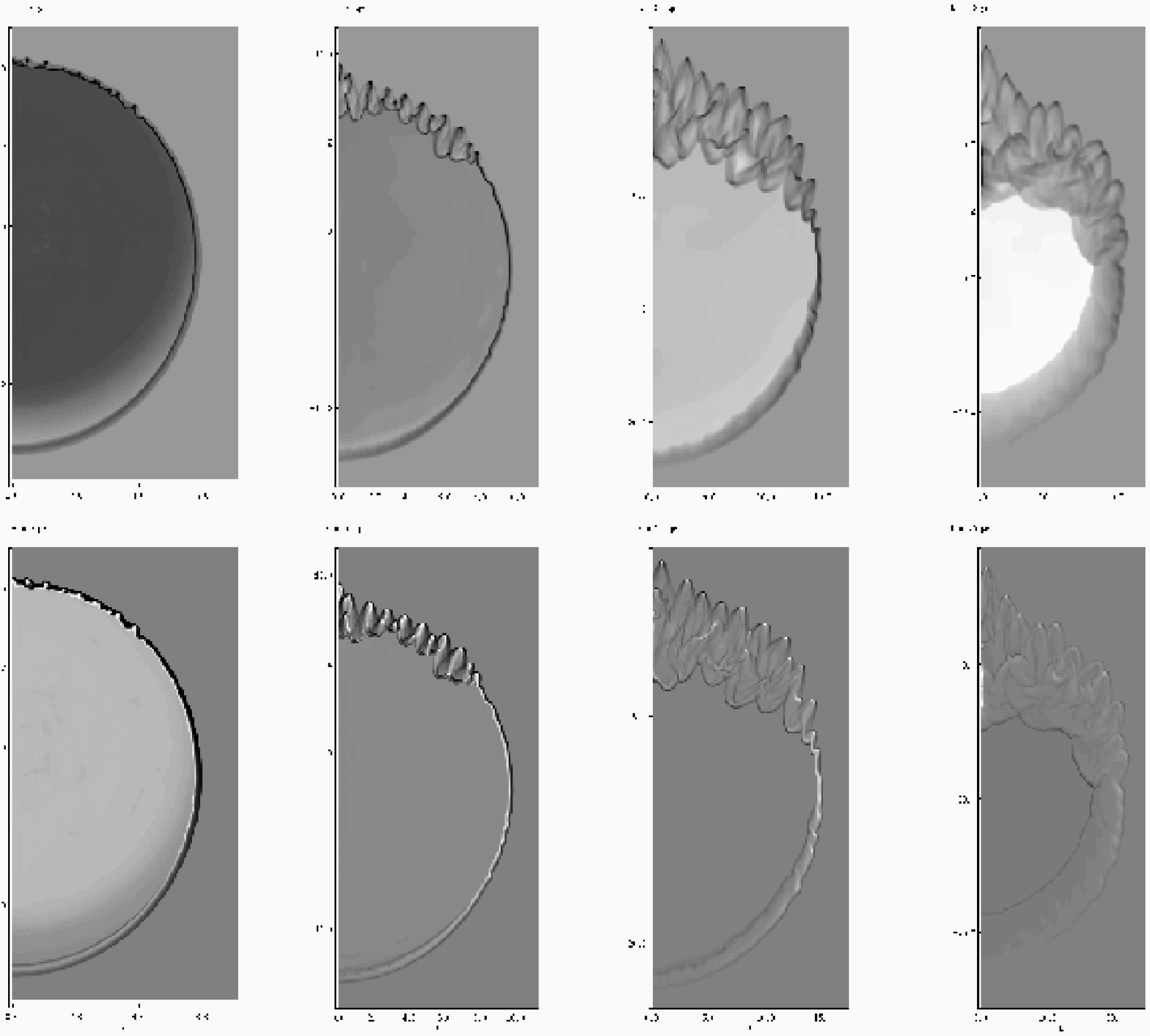,width=17cm}
\end{center}
\caption[]{The evolution of a SNR expanding into an AGN wind with 
$n = 10^{6} \pcm3$ and $v = 3000 \kmps$. In this figure the AGN wind propagates
down the page and the explosion site is at position (0,0). The top row shows 
logarithmic plots 
of the density with a scale from $2.4 \times 10^{4} \pcm3$ (white) to 
$3.1 \times 10^{8} \pcm3$ (black), while the bottom row shows logarithmic 
plots of the temperature with a scale from $10^{6} \K$ (white) to 
$1.58 \times 10^{8} \K$ (black). The age of the remnant is shown in the upper 
left corner of each plot. Note the change in the axis scaling between 
the plots.}
\label{fig:nw6_w3e8}
\end{figure*}

Several major differences in the morphology and evolution of the remnant
are seen when it expands into a high velocity flow rather than into a 
stationary medium. Fig.~\ref{fig:nw6_w3e8} shows density and temperature 
plots of a remnant expanding into a flow with $n = 10^{6} \pcm3$ and 
$v = 3000 \kmps$. Since the expansion velocity of the remnant is intially
much higher than this, the flow velocity of the surrounding medium has little
effect on the remnant morphology at early times. However, as the expansion 
velocity of the remnant slows, the flow velocity becomes increasingly
significant, and the remnant is distorted from a spherical shape in the way
that one would expect. The resulting morphology can be compared to that seen
in adiabatic remnants expanding into a plane-stratified (stationary) 
medium (Arthur \& Falle \cite{AF1993}). The models show increasing distortion 
of the remnant with wind speed.

Motion of the ambient medium also affects the development of hydrodynamic 
instabilities. On the leading edge of the remnant,
instabilities are more vigorous, since the
shocked region is compressed to higher densities. On the trailing edge, the shocks
are much weaker, and the shocked gas is both less dense and broader in extent, 
which severely suppresses the activity of the instabilities in this region. 
Since the trailing
edge has relatively low pressure, $\Xi$ is much higher here than at the leading 
edge, and cool gas forms preferentially in the upstream direction. 
We find that the evolution of the mass of cool gas is surprisingly 
similiar over a wide range of ambient densities and flow speeds 
(see Sec.~\ref{sec:cool_gas}).

\subsection{Individual cool clumps}
\label{sec:individual_clumps}
\begin{figure*}[t]
\begin{center}
\psfig{figure=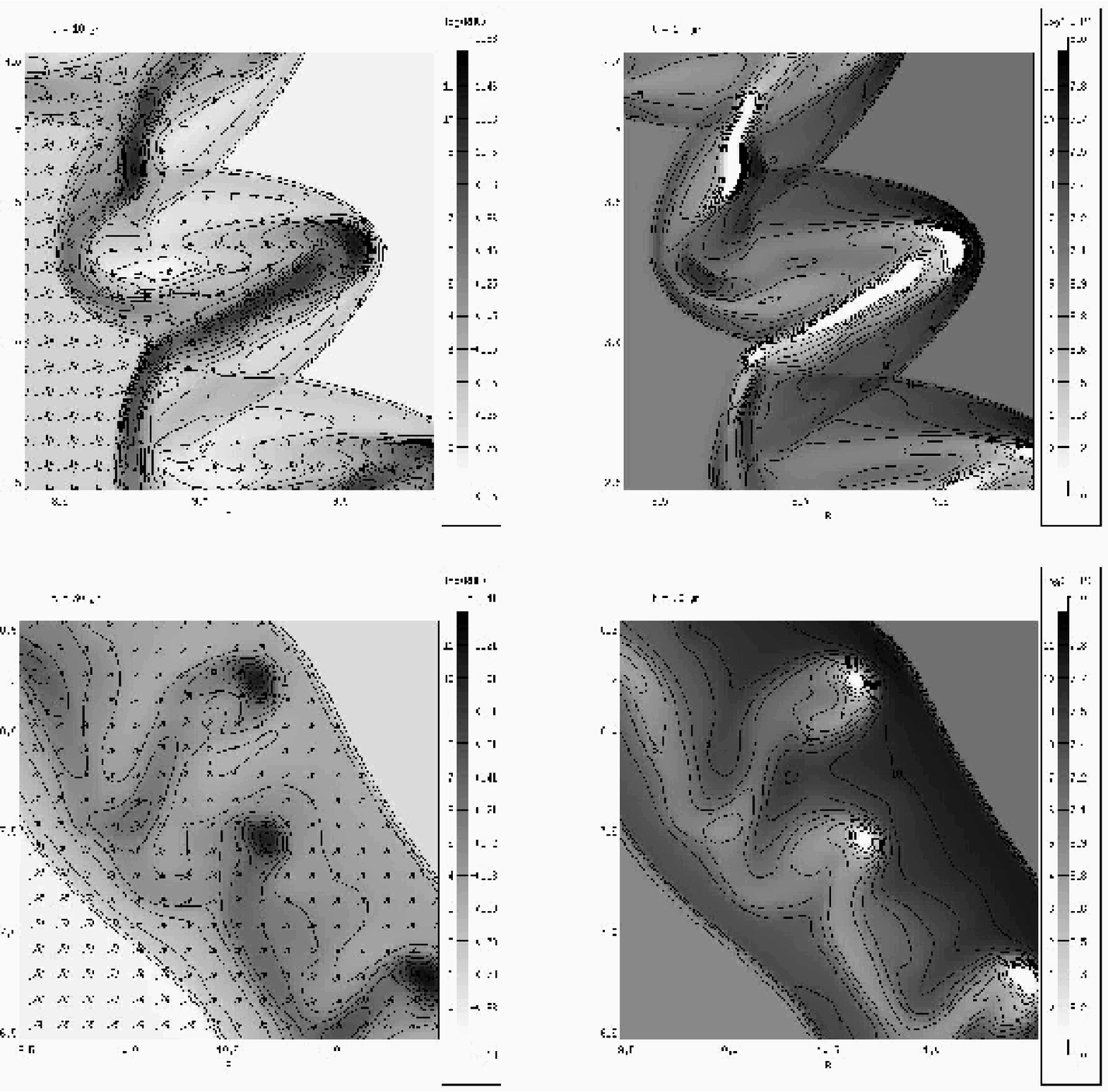,width=17cm}
\end{center}
\caption[]{The density and temperature structure of individual bowshocks 
and cool clumps. The top panels show the structure existing at $t = 10 \yr$ when a 
remnant expands into an ambient density $n = 10^{6} \pcm3$. The bottom 
panels show the structure existing at $t = 30 \yr$ when a 
remnant expands into an ambient density $n = 10^{5} \pcm3$. The left panels
display density gray-scales, contours, and flow vectors, while the right
panels display temperature gray-scales and contours.}
\label{fig:clump}
\end{figure*}

In the upper panels of Fig.~\ref{fig:clump} we zoom in on the remnant limb 
from the simulation at $t = 10 \yr$ as shown in Figs.~\ref{fig:nw6_rho} 
and~\ref{fig:nw6_temp}. The forward shock bounds the right side 
of the shocked gas, and the reverse shock bounds the left. The high densities
apparent in the upper panels lead to rapid cooling of the post-shock flow,
such that the morphology resembles that resulting from the ram-ram pressure 
instability. The lower panels of Fig.~\ref{fig:clump} display the remnant limb
at $t = 30 \yr$ for expansion into an ambient density $n = 10^{5} \pcm3$ 
(see Fig.~\ref{fig:nw5}). Compared to the situation shown in the upper 
panels the post-shock flow cools far less strongly, and we see that the cool 
regions are embedded in a much thicker region of shocked gas. 

Of note is the fact that the cool regions (which we shall identify 
henceforth as ``clouds'') contain within them a range of densities,
equilibrium ionization parameters, and velocities. The clouds, being 
surrounded by a confining medium, are indeed long-lived: they remain as 
distinct entities until their pressure drops to the point where their
equilibrium temperature corresponds to the hot phase. Within an AGN 
as a whole where many young SNRs will exist at a given time, cool clouds
will be continuously created (through gas cooling in the younger remnants) 
and destroyed (through gas re-heating in the older remnants).

\subsection{Evolution of cool gas}
\label{sec:cool_gas}

\begin{figure*}[t]
\begin{center}
\psfig{figure=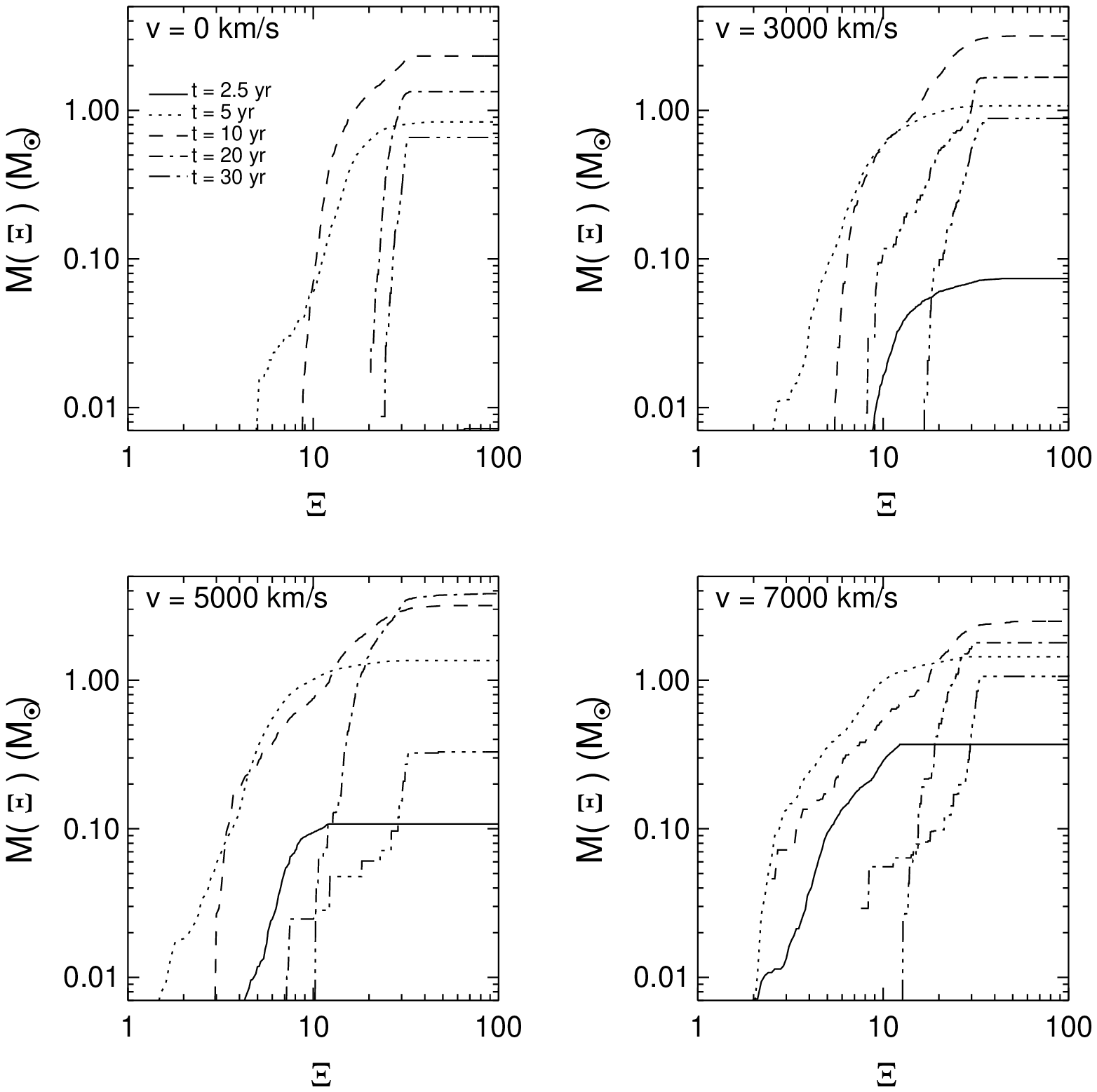,width=14.5cm}
\end{center}
\caption[]{The evolution of cool gas in remnants expanding into a surrounding
medium with $n = 10^{6} \pcm3$. The mass of cool gas with an ionization parameter
below $\Xi$ is shown as a function of $t$ for flow speeds of 
$v = 0,\;3000,\;5000,\;7000 \kmps$.}
\label{fig:mass_ip_evolution}
\end{figure*}

\begin{figure}
\begin{center}
\psfig{figure=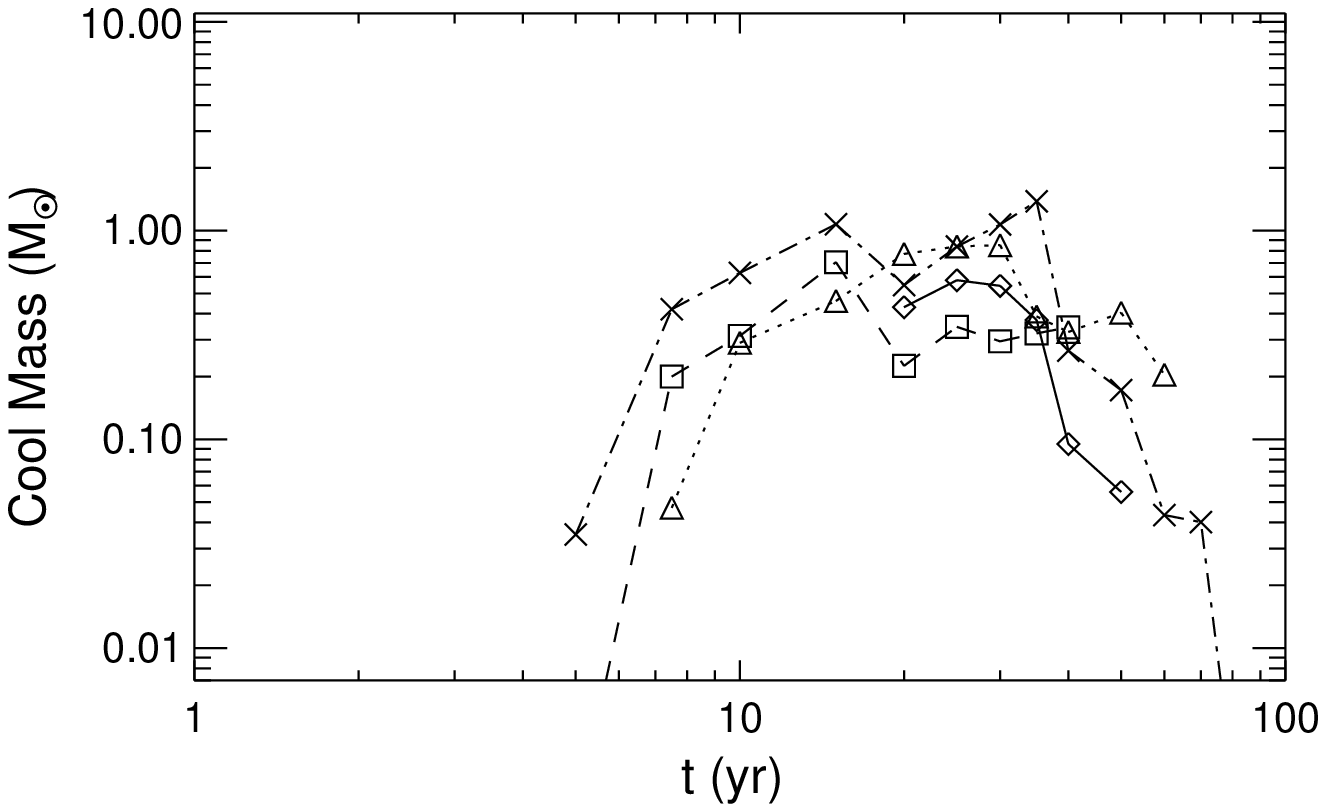,width=8.8cm}
\psfig{figure=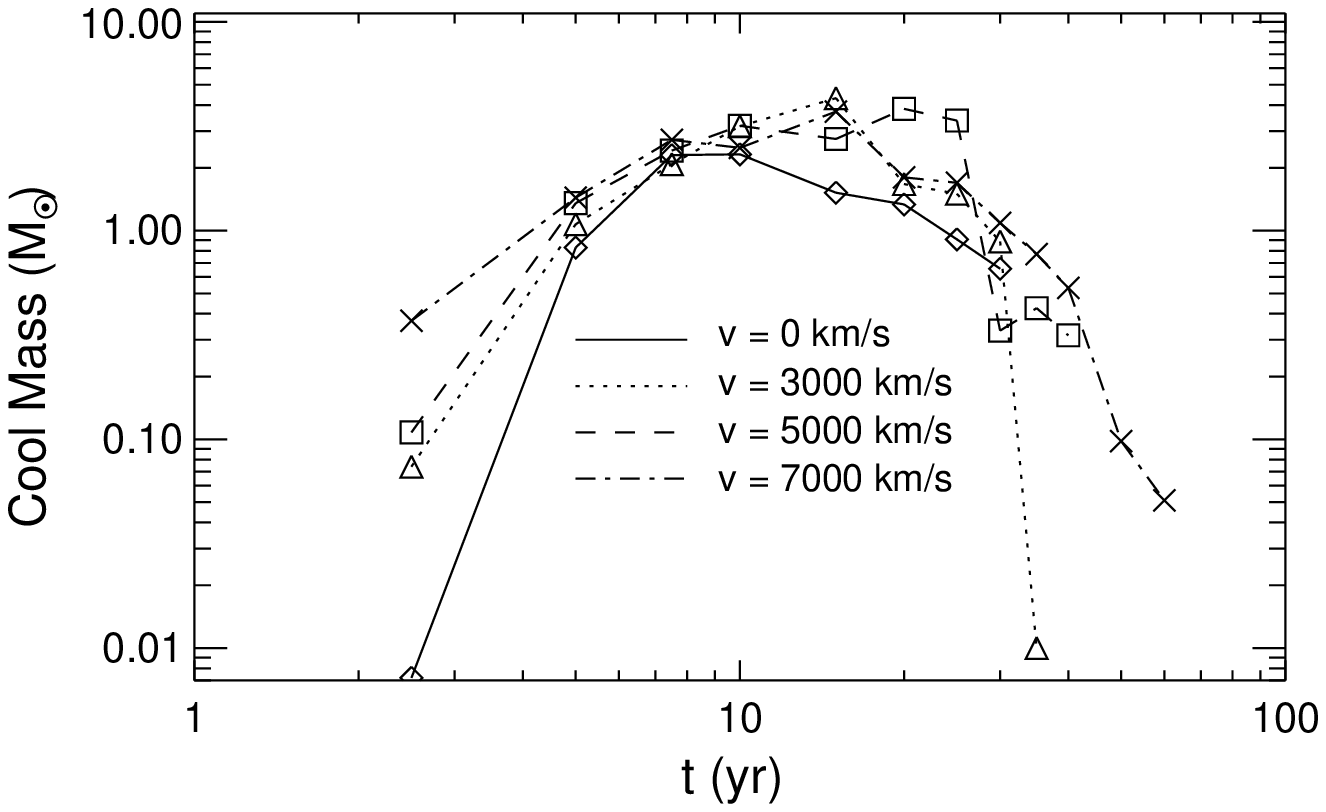,width=8.8cm}
\end{center}
\caption[]{The evolution of the mass of cool clouds in remnants expanding 
into a surrounding medium with $n = 10^{5} \pcm3$ (top) and 
$n = 10^{6} \pcm3$ (bottom). 
In each panel the mass of cool gas is shown as a function
of $t$ for flow speeds of $v = 0,\;3000,\;5000,\;7000 \kmps$.}
\label{fig:mass_time_evolution}
\end{figure}

The mass of the cool clouds as a function of time is shown in 
Figs.~\ref{fig:mass_ip_evolution} and~\ref{fig:mass_time_evolution},
and Table~\ref{tab:mass_cool_gas}. We define 
$M = \int_{\Xi = 0}^{\Xi} M(\Xi) d\Xi$.
The first gas to cool below $T = 10^{6} \K$ is initially out of pressure
equilibrium with its surroundings as the cooling timescale of this gas 
is shorter than the dynamical timescale of the surrounding material. This
can be seen by examining Fig.~\ref{fig:mass_ip_evolution}, where during  
the re-establishment of pressure equilibrium, $\Xi$ decreases. 
For the remaining evolution, continued expansion of the 
remnant causes $\Xi$ to increase towards the ambient value ($\Xi \approx 150$).
While $\Xi$ is below the value separating the cool and hot phases
($\Xi \approx 30$ - see Fig.~\ref{fig:teq}), gas continues to cool and the
total mass of cool gas continues to rise. However, when $\Xi \gtsimm 30$,
the clouds are subject to net heating, and the mass of cold
gas decreases until the clouds are eventually completely destroyed. 
Fig.~\ref{fig:mass_ip_evolution} shows that
not all of the clouds transit from a cool to a hot phase at exactly 
the same time. Pressure and density enhancements in the remnant caused by
the action of instabilities are able to maintain some clouds in their cool 
phase for a longer duration.

When the surrounding medium is in motion relative to the progenitor of the
explosion, the increased compression of the leading edge of the remnant
results in cool gas forming more rapidly and with a lower value of $\Xi$.
The cool gas is also able to survive for a longer time 
in such cases. For $n = 10^{6} \pcm3$, 
the total mass of cool gas is fairly insensitive to the flow speed 
of the ambient medium, peaking at $M \sim 2-4 \Msol$ approximately $15-20 \yr$ 
after the explosion. On the other hand, the increased compression caused 
by a wind can significantly enhance the mass of cool gas in situations where
its formation is otherwise marginal. For example, when $n = 10^{5} \pcm3$, 
gas barely manages to cool below $T \approx 10^{6} \K$ when the medium is 
stationary, but in the model with $n = 10^{5} \pcm3$ and 
$v = 3000 \kmps$ a substantially greater mass of gas cools and exists for
a significantly longer period.

The results of most observational work are reported in terms of the 
ionization parameter $U$ (Eq.~\ref{eq:u_ip}). Since we know the relationship
between $U$ and $\Xi$ for the AGN spectrum adopted in our models 
(Eq.~\ref{eq:u_chi}), we can therefore easily compare our results 
to observations, where a large range in $U$ is seen. Low ionization lines    
(such as \ion{Mg}{ii} $\lambda2800$) typically have $U \sim 0.01$, which
translates into $\Xi \sim 0.3$ for gas at $T \approx 20000 \K$. In
contrast, high ionization lines (such as \ion{C}{iv} $\lambda1549$) 
are characterized by larger values of $U$ - low values of 
\ion{C}{iii} $\lambda977$/\ion{O}{vi} $\lambda1034$ imply the presence
of gas with $U \sim 1$ (Laor \etal \cite{L1994}). This translates into
$\Xi \sim 25$ for gas existing in the cool phase, which is clearly
in good agreement with the results presented in 
Fig.~\ref{fig:mass_ip_evolution}.

\begin{table*}
\begin{center}
\caption{Time evolution of the mass ($\Msol$) of cool clouds in remnants expanding into
a surrounding medium with various density, flow speed, and ionization parameter. 
Columns 2-5 show results for an ambient density $n = 10^{5} \pcm3$.
Columns 6-11 are for an ambient density $n = 10^{6} \pcm3$. 
In columns 10 and 11 we give results for stationary surroundings
but with either an enhanced AGN flux (causing a 10x increase in the 
ionization parameter of the ambient medium to $\Xi \approx 1500$), or 
remnant parameters suitable for a type Ia SN explosion 
(see Secs.~\ref{sec:high_agn_flux} and~\ref{sec:stellar_envelope} respectively).
Dashes indicate that no cool gas exists. The bottom row contains values for the
integrated area under the mass vs. time curve, in units of $\Msol \yr$.}
\label{tab:mass_cool_gas}
\begin{tabular}{|l|l|l|l|l|l|l|l|l|l|l|}
\hline
$t (\yr)$ & \multicolumn{4}{|c|}{$n = 10^{5} \pcm3$} & \multicolumn{6}{|c|}{$n = 10^{6} \pcm3$} \\
\hline
 & 0 & 3000 & 5000 & 7000 & 0 & 3000 & 5000 & 7000 & High $\Xi$ & SN~Ia \\
\hline
2.5 & - & - & - & - & 0.0072 & 0.074 & 0.11 & 0.37 & - & - \\
5.0 & - & - & 0.0015 & 0.035 & 0.83 & 1.1 & 1.4 & 1.5 & - & - \\
7.5 & - & 0.047 & 0.20 & 0.42 & 2.3 & 2.1 & 2.4 & 2.7 & - & 0.51  \\
10.0 & - & 0.29 & 0.31 & 0.63 & 2.3 & 3.2 & 3.2 & 2.5 & - & 0.79  \\
15.0 & - & 0.46 & 0.71 & 1.1 & 1.5 & 4.3 & 2.7 & 3.7 & - & 0.30  \\
20.0 & 0.48 & 0.77 & 0.23 & 0.55 & 1.3 & 1.7 & 3.8 & 1.8 & - & 0.033 \\
25.0 & 0.58 & 0.84 & 0.35 & 0.84 & 0.91 & 1.5 & 3.4 & 1.7 & - & - \\
30.0 & 0.54 & 0.85 & 0.30 & 1.1 & 0.66 & 0.89 & 0.33 & 1.09 & - & -  \\
35.0 & 0.37 & 0.39 & 0.32 & 1.4 & - & 0.01 & 0.43 & 0.77 & - & -  \\
40.0 & 0.095 & 0.33 & 0.34 & 0.27 & - & - & 0.32 & 0.53 & - & -  \\ 
50.0 & 0.056 & 0.40 & - & 0.17 & - & - & - & 0.098 & - & - \\
60.0 & - & 0.20 & - & 0.043 & - & - & - & 0.051 & - & -  \\
70.0 & - & - & - & 0.040 & - & - & - & - & - & - \\
80.0 & - & - & - & 0.0025 & - & - & - & - & - & - \\
90.0 & - & - & - & - & - & - & - & - & - & -  \\
\hline
$Mt$ & 11 & 26 & 14 & 33& 39 & 62 & 77 & 71 & - & 6 \\
\hline
\end{tabular}
\end{center}
\end{table*}

\subsection{Effect of an increased AGN flux}
\label{sec:high_agn_flux}
\begin{figure*}[t]
\begin{center}
\psfig{figure=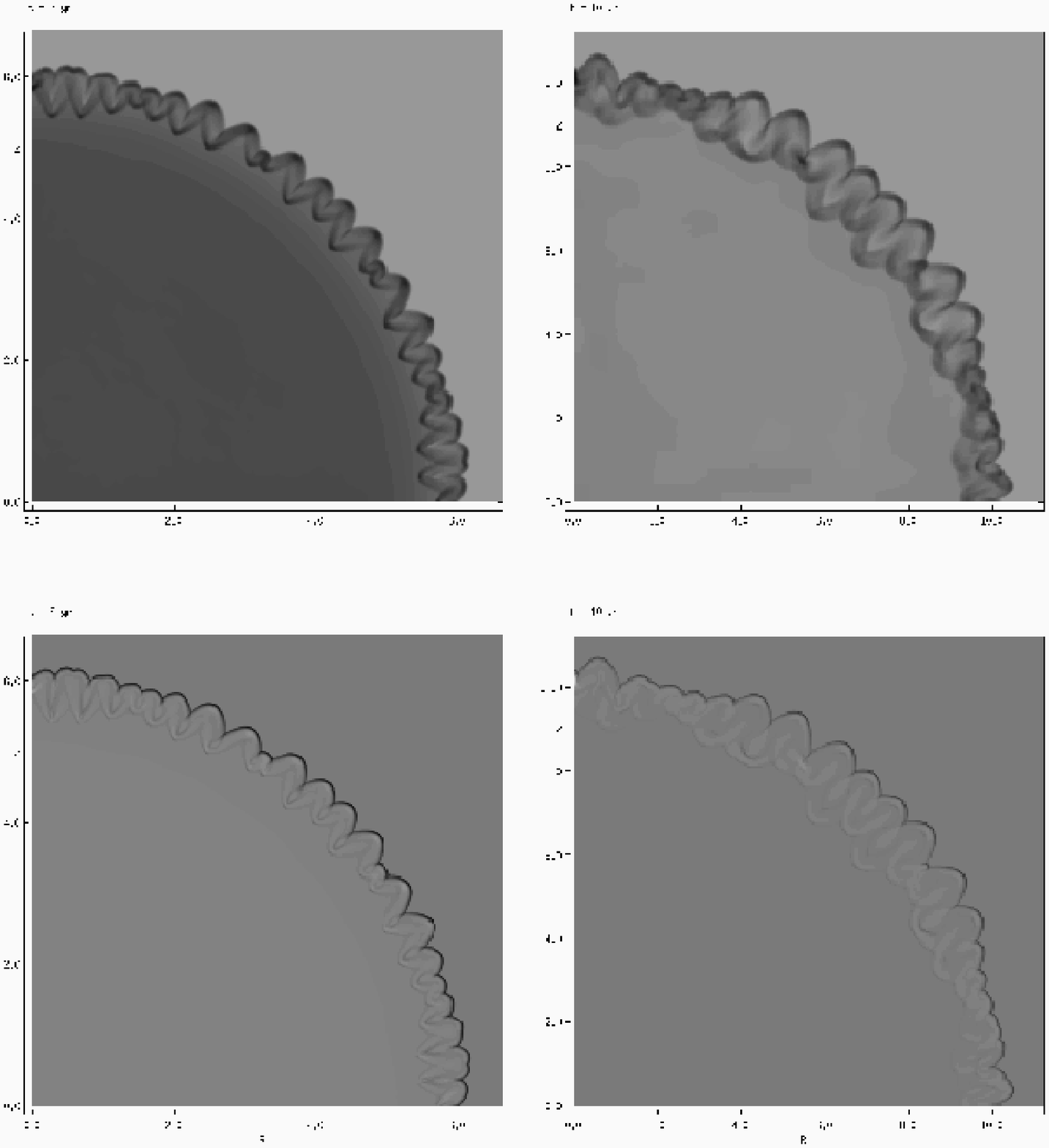,width=17cm}
\vspace{-15mm}
\end{center}
\caption[]{The evolution of a SNR expanding into a stationary environment 
with $n = 10^{6} \pcm3$ and $\Xi \approx 1500$. The top panels show 
logarithmic plots of the density, while the bottom panels show logarithmic 
plots of the temperature (same scaling as in Figs.~\ref{fig:nw6_rho} 
and~\ref{fig:nw6_temp}). Note that cool clouds do not form in this case.}
\label{fig:nw6_highip}
\end{figure*}

The simulations shown so far were computed with an AGN flux which was high 
enough to maintain the ambient gas in the hot phase  (\ie $\Xi \approx 150$ - see 
the equilibrium curve in Fig.~\ref{fig:teq}). This resulted in a 
relatively low value of $\Xi$ for the shocked gas and provided a good
opportunity for the shocked gas to cool to $T \approx 2 \times 10^{4} \K$.
However, if a remnant were to be immersed in ambient gas with a higher ionization 
parameter, the ability for shocked gas to cool would be compromised. This is shown in 
Fig.~\ref{fig:nw6_highip}, where a remnant is expanding into a stationary medium
with $\Xi \approx 1500$ (\ie $\Xi$ is 10x higher than in Figs.~\ref{fig:nw6_rho} 
and~\ref{fig:nw6_temp}). The increased photon flux from the AGN increases the 
rate of Compton heating, and leads to a reduction in the net cooling rate. 
While the radius of the remnant is largely unchanged, its morphology is
affected: the shocked gas is not compressed as much, and cool
clouds do not form. This behaviour is consistent with the earlier work presented in 
Pittard \etal (\cite{PDFH2001}).

\subsection{Effect of the stellar envelope}
\label{sec:stellar_envelope}
To investigate whether cool clouds could form in a remnant from a 
type Ia SN explosion, we have computed an additional model with appropriate
parameters: $E = 10^{51} \erg$, $M = 1.4 \Msol$, and $n = 7$ (\cf 
Chevalier \cite{C1982}). The main difference with this model is that 
the increased value of $E/M$ results in higher expansion 
velocities of the ejecta and hotter post-shock gas, and a lower 
equilibrium ionization parameter. However, the remnant also decelerates
more rapidly, causing $\Xi$ to increase at a faster rate than for the 
$n = 12$ case. This ultimately hinders the formation of cool gas relative to
the $n = 12$ case, and only a small amount is able to form 
(see Table~\ref{tab:mass_cool_gas}). 

Recently, it is has become clear that there exists a class of type II
supernova explosions which are under energetic (\eg, Zampieri \etal \cite{Z2003}).
For two explosions examined in detail, Zampieri \etal (\cite{Z2003}) found
$M \gtsimm 14 \Msol$ and $E \approx 0.6 - 0.9 \times 10^{51} \erg$.
We do not expect remnants with such parameters to evolve significantly 
differently to our canonical models with $M = 10 \Msol$ and $E = 10^{51} \erg$.

\section{Line profiles}
\label{sec:line_profiles}
We have calculated synthetic line profiles for emission from the cooled gas.
The 2D axisymmetric grid is rotated onto a 3D cartesian grid and the emission 
from volume elements containing cool gas was integrated under the assumption that it 
is optically thin and the volume emission rate varies as $n^{2}$. 
Since the gas is cool, thermal Doppler broadening is negligible. A previous
investigation (Bottorff \etal \cite{BFBK2000}) failed to reach any strong
conclusions concerning whether microturbulence\footnote{Defined as any velocity
field that occurs over distances that are small compared to a photon's mean
free path.} was favoured by observations, so it is not included in our model.
The emission is blue- or redshifted according to the line of sight
velocity of the gas. Absorption was also assumed to be negligible.

\begin{figure*}[t]
\begin{center}
\psfig{figure=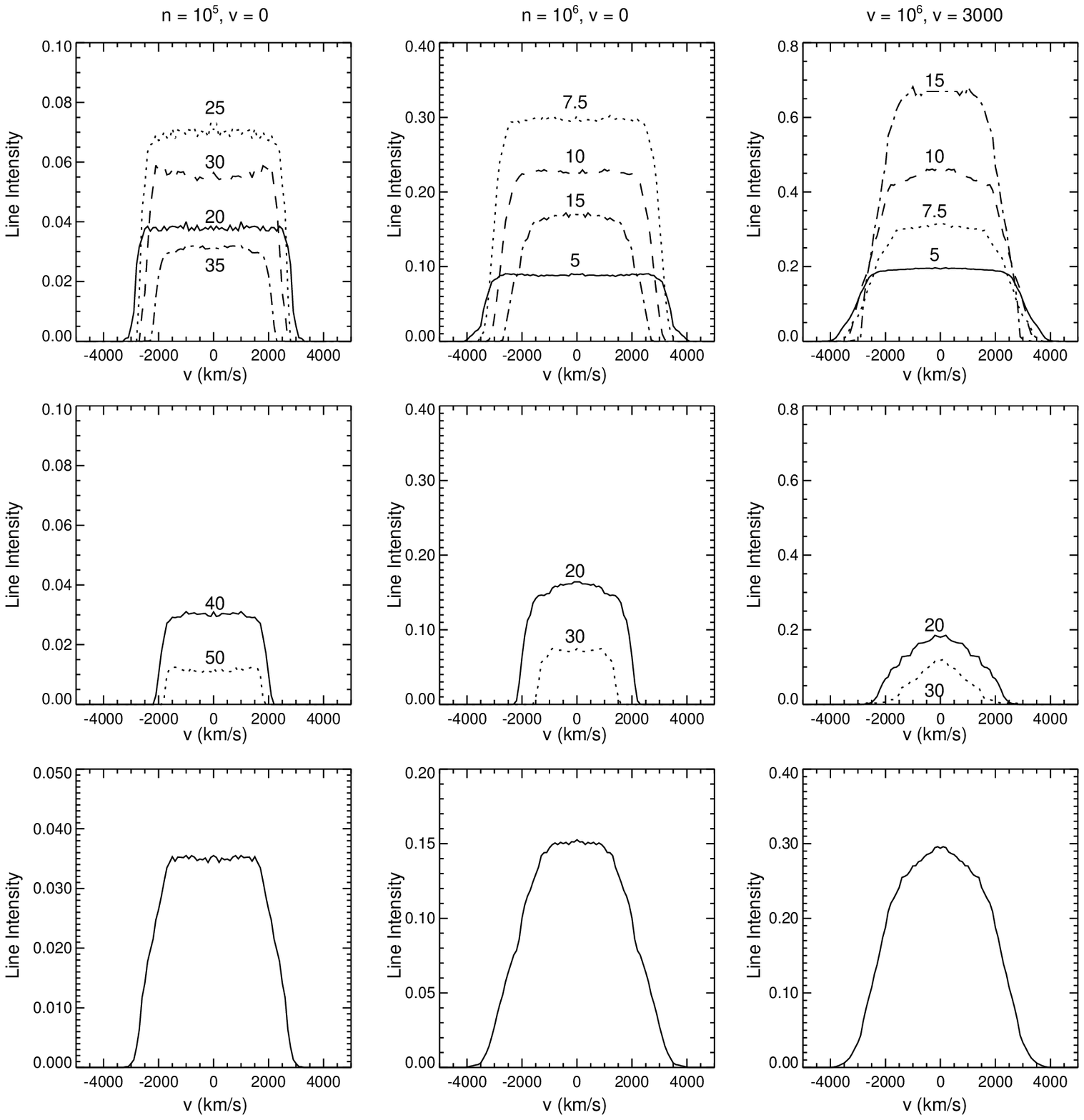,width=14.5cm}
\end{center}
\caption[]{Line profiles from cool gas formed in a SNR expanding into an
AGN environment with $\Xi \approx 150$ and integrated over many lines of
sight (\ie it is assumed that there is a spherical distribution of remnants
in an isotropic medium and radiation field). The left and middle columns are 
for stationary environments with $n = 10^{5} \pcm3$ and $n = 10^{6} \pcm3$,
respectively. The right column is for an environment with $n = 10^{6} \pcm3$
and $v = 3000 \kmps$. In the top two rows the line profile at specific 
remnant ages is displayed, with the age of the remnant (in years) marked 
next to each profile. The solid line in the top row corresponds to
$t = 20\yr$ for $n = 10^{5} \pcm3$, and $t = 5\yr$  for $n = 10^{6} \pcm3$.
The remaining profiles in the top and middle rows correspond to the subsequent 
entries in Table~\ref{tab:mass_cool_gas}, with the line width decreasing with 
remnant age. In the bottom row the average line profile integrated over the age 
of the remnant is displayed. Note the difference in the scaling of the 
intensity axis.}
\label{fig:lp1}
\end{figure*}

In Fig.~\ref{fig:lp1} we show the line profiles resulting from a remnant
expanding into an ambient medium with 3 sets of parameters. 
The normalization of the line profiles has been set so that the peak 
emission is approximately 1.0, and we have preserved the relative scaling 
between the models (\eg, the central intensity of the $n = 10^{6} \pcm3$,
$v = 0 \kmps$, $t = 20\yr$ profile is approximately $4\times$ greater than
the central intensity of the $n = 10^{5} \pcm3$,
$v = 0 \kmps$, $t = 20\yr$ profile). The bottom row in Fig.~\ref{fig:lp1} 
shows the line profile which results if we integrate over the age of 
the remnant. In effect we sum each of the profiles in the rows above 
with an appropriate weight which reflects the time between each ``snapshot''.

In general, the 
normalization first rises and then falls with time as cool clouds
are created and then destroyed. The width of the line decreases with time
as the expansion speed of the remnant slows. The line is clearly flat-topped,
which is characteristic of emission from a geometrically and 
optically thin spherical shell (see, \eg, Fig.~3 in Capriotti \etal 
\cite{CFB1980} with $v_{\rm min}/v_{\rm max} \approx 1.0$). As the effective 
thickness of the ``shell'' in Fig.~\ref{fig:nw6_rho} is minimal, a tangential 
line of sight does not intercept an increased number of clouds, and 
``horns'' are not seen in the profile. In contrast, the line profiles
displayed in the rightmost column of Fig.~\ref{fig:lp1} show an increasingly
rounded or triangular profile as the remnant ages, and the time-averaged
profile displays a distintly rounded top. In this model the
remnant is expanding into an AGN wind and the line profiles (which are sensitive
to the spatial distribution of cool gas) reflect the increasing distortion
of the remnant as it expands.

It is clear from Fig.~\ref{fig:nw6_rho} that the emission is from a relatively 
small number of clouds (particularly at later times, \eg, $t = 20 \yr$).
This introduces a great deal of small scale structure into the line profiles
which we have smoothed out in Fig.~\ref{fig:lp1} by averaging over
many different lines of sight. Observed line profiles from AGN are in 
reality very smooth, and it has been concluded that the number of
emitting clouds must be $\sim 10^{5}$ (Arav \etal \cite{A1997}),
although this would be reduced if there was significant microturbulence.
Alternatively, electron scattering could help to explain smooth broad
line profiles, especially in the line wings (Emmering \etal \cite{EBS1992}).  In our
2D hydrodynamical models the clouds are in fact rings, and numerical viscosity and
the finite number of grid cells limit the number of distinct clouds which form.
Increased numerical resolution and 3D simulations will produce many 
more distinct clouds, but present limitations mean that the line 
profiles from our models are not as smooth as seen in observations, 
and their fine structure should be ignored. 
 
\begin{figure*}[t]
\begin{center}
\psfig{figure=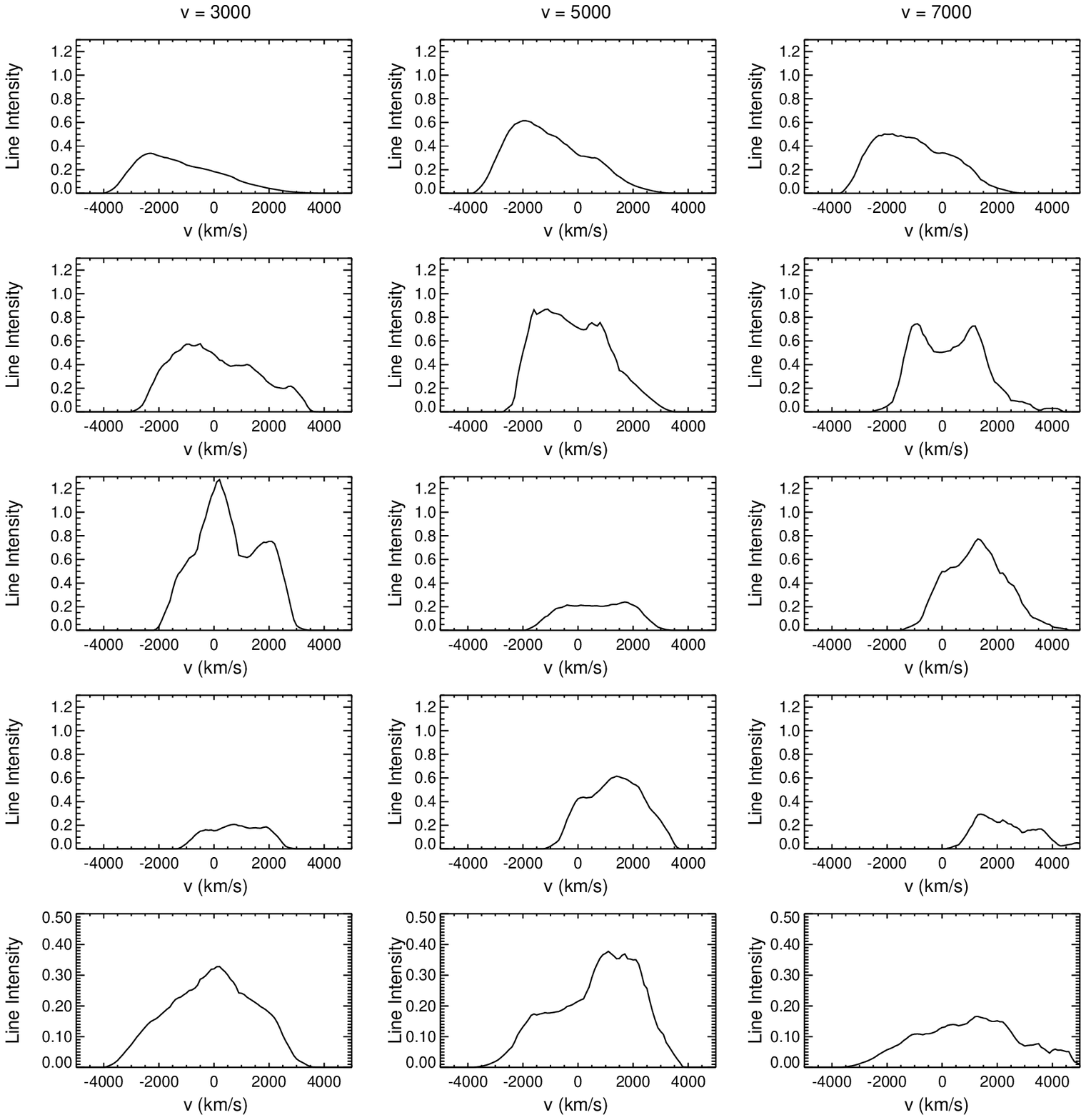,width=14.0cm}
\end{center}
\caption[]{Line profiles from cool gas formed in a SNR expanding into an AGN 
wind with $n = 10^{6} \pcm3$, $\Xi \approx 150$, and $v = 3000, 5000,$ or 
$7000 \kmps$. The line-of-sight is specified by $\theta = 135^{\circ}$ (\ie there 
is a component of the AGN wind moving towards the observer) 
and the profiles are from remnants of age 
$t = 5\yr$ (top row), $t=10\yr$, $t=15\yr$, and $t=20\yr$. In the bottom row
the age integrated profiles are shown.}
\label{fig:lp2}
\end{figure*}

In Fig.~\ref{fig:lp2} we show the line profiles resulting 
from a remnant expanding into an AGN wind of speed $v = 3000, 5000$, and 
$7000 \kmps$, for a viewing angle, $\theta = 135^{\circ}$. Here $\theta$
is defined as the angle between the AGN wind vector and the vector from the observer
to the remnant (\ie $\theta = 0^{\circ}$ corresponds to the observer
facing the side of the remnant expanding into the oncoming AGN wind). 
Inspection of Fig.~\ref{fig:nw6_w3e8} reveals that
cool clouds exist over a wide range of angles, namely from $\theta = 
0^{\circ}$ to $\theta \approx 135^{\circ}$. For a line of sight with 
$\theta = 135^{\circ}$ (\ie where the flow has a velocity component 
towards the observer), the
blue wing of the line shows the greater extension at $t = 10 \yr$. 
This is expected since the trailing edge of the remnant is not decelerated 
as rapidly as the leading edge. However, at $t= 5 \yr$, the cold clouds
exist predominantly on the leading edge (the density and cooling rate are
highest here) and it is the red edge of the line profile 
that is the most extended. This behaviour is also seen when the
remnant expands into a flow with $v = 5000$ or $7000 \kmps$.
Hence as the remnant expands, the resulting line profile flips from
a red to a blue-shift (and vice-versa for $\theta \ltsimm 45^{\circ}$).

For each of the 3 distinct cases of ambient medium investigated in 
Fig.~\ref{fig:lp1}, the time-averaged line profile displayed in the 
bottom row has a FWHM~$\approx 4500-5000 \kmps$, and
a full width at zero intensity of $\approx 6000-8000 \kmps$.
This range lies almost in the middle of the 
distribution of FWHM measures of the broad component of 
\ion{C}{iv} $\lambda1549$ for radio-quiet sources (Sulentic \etal 
\cite{SMD2000}). 

In Table~\ref{tab:line_params} we list 
various statistics for the time-averaged line profiles shown in 
Fig.~\ref{fig:lp2}. Radio quiet sources show a scatter in the 
asymmetry values for the broad component of the 
\ion{C}{iv}$\lambda 1549$, with ${\rm A.I.} = \pm 0.2$, and
$\langle {\rm A.I.} \rangle \approx +0.04$ (Marziani \etal \cite{M1996}).
This lack of a strong preference for red or blue asymmetries is
currently much in need of confirmation, especially when one 
considers the systematic blueshift that \ion{C}{iv}$\lambda 1549$
shows in RQ sources (see Sulentic \etal \cite{SMD2000} and references
therein). It is interesting to note that our models produce values
for the asymmetry index which are compatible with the observationally
deteremined values. 

\begin{table}
\begin{center}
\caption{Statistics for the time-averaged line profiles shown in 
Fig.~\ref{fig:lp2} ($n = 10^{6} \pcm3$). Detailed are the profile
centroid measures at different levels of the line peak, the 
full-width at half maximum (FWHM), and the asymmetry index, 
${\rm A.I.} = ({\rm C}({3\over4})-{\rm C}({1\over4}))/{\rm FWHM}$.
The columns labelled $v_{3000}$, $v_{5000}$, and $v_{7000}$ refer
to the models where the AGN wind speed is $3000, 5000$ and 
$7000 \kmps$ respectively.}
\label{tab:line_params}
\begin{tabular}{llll}
\hline
\hline
Statistic & $v_{3000}$ & $v_{5000}$ & $v_{7000}$ \\
\hline
C(1) ($\kmps$)   & 0 & 900 & 1000 \\
C(3/4) ($\kmps$) & 0 & 1400 & 1200 \\
C(1/4) ($\kmps$) & -100 & 500 & 1200 \\
FWHM ($\kmps$)   & 4200 & 3200 & 4200 \\
A.I.             & 0.0 & 0.3 & 0.0 \\
\hline
\end{tabular}
\end{center}
\end{table}

Observations have revealed great diversity in the profiles of particular
broad emission lines, such as \ion{C}{iv} $\lambda1549$, whereas variations in other
lines, such as \ion{O}{iv]} $\lambda1402$, are much reduced in comparison 
(see, \eg, Francis \etal \cite{FHFC1992}; Wills \etal \cite{W1993}). 
The variation in \ion{C}{iv} $\lambda1549$ from object to object can be 
explained if a two-component structure is invoked - a broad ``base'' which
is relatively constant between objects, and a narrow ``core'' whose 
contribution to the overall line strength is variable. With this model, lines with
large equivalent widths tend to be narrow (or ``cuspy'') and have large
peak fluxes, as observations require. In contrast, 
lines such as \ion{O}{iv]} $\lambda1402$,
which show far less variability from object to object, are essentially
composed of a single broad component. In such cases, the contribution from
a variable narrow core is small.

The observational interpretation of broad line profiles has advanced
greatly over the last decade, partly due to a better understanding of sample
biases and more careful consideration of contaminating lines and superposed
narrow components. Sometimes an obvious inflection demonstrates a 
superposed narrow component. Alternatively, if a line has a rounded top, 
one can infer that the narrow component is largely absent. 

Inflections also support the interpretation that the 
broad lines are formed in several kinematically and/or geometrically
distinct emitting regions, as does the frequent ``mismatch'' in profile
wings (Romano \etal \cite{RZCS1996}).
One possibility is that the variable narrow component forms on the outer
edge of the BELR (this is sometimes referred to as an ``intermediate line 
region''), with the broad component formed at smaller radii within the 
BELR. Variations are then most likely explained as a changing covering 
factor (if optically thick) or volume emissivity at the outer edge. 
Support for this model comes from variability measurements of
NGC~5548 and NGC~4151 (Clavel \cite{C1991b}), in which the line cores
(defined as the central $3000 \kmps$) lag continuum changes with approximately 
twice the delay of the wings. An alternative model consisting of a 
biconical outflow and an accretion disc continuum source has been
shown to have problems in explaining the near-constant velocity width
and equivalent width of the emission line wings (Francis \etal \cite{FHFC1992}),
and there are doubts concerning the dependence of the line equivalent
width on orientation via. an axisymmetric continuum, which this model 
requires (Wills \etal \cite{W1993}).

\section{Discussion}
\label{sec:conclusions}
The results in Sec.~\ref{sec:results} demonstrate that it is possible to
cool shocked supernova ejecta to $T \sim 10^{4}$~K in the inner 
regions of a QSO. Although our results differ from the original proposals 
of Dyson \& Perry (\cite{DP1982}) and Perry \& Dyson (\cite{PD1985}), which
were for the shocked ambient medium to cool, the resulting cool
gas nevertheless has properties (densities, column densities, velocities and 
ionization parameters) compatible with those inferred for gas 
emitting the high ionization lines in QSOs. 

A parameter space study shows that for ambient densities of
$n = 10^{5} - 10^{6} \pcm3$, about $1-4 \Msol$ of material can cool
to low temperatures. Such gas then persists for $\sim 10-20 \yr$ before 
the continuing expansion of the remnant reduces the density and 
pressure of the cool gas to the point where its equilibrium temperature 
and ionization parameter corresponds to the hot phase. This result
is robust for a wide range of velocities of the surrounding medium, although
the spatial distribution of the cool gas around the limb of the remnant,
and hence the resulting optical/UV line profile, is sensitive to this 
detail. We find that the integrated value of the mass of cool gas over the
remnant lifetime generally increases with the density and the flow 
velocity of the surrounding medium. The highest value found in our simulations,
$Mt = 77 \Msol \yr$, is obtained when $n = 10^{6} \pcm3$ and $v = 5000 \kmps$.

A supernova rate of 1/yr would then imply a mass for the clouds emitting
the HILs of up to $\sim 80 \Msol$,
This is easily compatible with the lower end of BELR mass estimates in the
literature (\eg, Peterson \cite{P1997}), although our model (and most others) would
be severely challenged to explain much more extreme estimates of 
the mass of BELR gas (see Baldwin \etal \cite{B2003} and references therein).
We note that it is currently unclear how this mass is partitioned between the 
HIL and LIL gas in these higher estimates.

In earlier work it was shown that for typical QSO parameters the power 
going into supernova remnants is comparable to that of the QSO wind, but
is much less than the bolometric QSO luminosity (Perry \& Dyson \cite{PD1985}).
However, it is more difficult to estimate whether emission from the
SN would be visible above the QSO in a specific waveband.
The typical J-band magnitude for a QSO at a redshift $z \sim 1$ is 
$\sim 18-19$, whereas the J+H band magnitude for a type Ia SN is $\sim 24$
at comparable $z$. On this basis, individual SN will not be 
discernible, but clearly this conclusion depends on the luminosity of the
QSO, as well as other variables such as the orientation of the SN with
respect to the molecular torus, and the ambient density of the
surroundings (\eg, if the SNR expands into a nearby molecular cloud then its
luminosity could be significantly increased). Detailed numerical
modelling will be required to determine the likelihood of this possibility.
 
One of the most interesting questions concerning AGNs is the connection
between nuclear and starburst activity.
Mixed starburst-AGN sources may be recognizable as outliers in the 
${\rm EW}(\ion{H}{\beta}) - {\rm EW}(\ion{Fe}{ii}\; \lambda4570)$ plane 
(Sulentic \cite{SMD2000}). In this sense AGNs displaying {\em galaxy-wide} 
starburst activity are in the minority. However, to provide enough
fuel for the high luminosity QSOs, the accreting medium needs to be 
particularly dense. Possible mechanisms to augment the density include 
the winds and explosions of massive stars, and galaxy
collisions. As the latter are often associated with starbursts, 
vigorous massive star formation in the inner regions of AGN is probably 
a necessity. If a significant component of the BELR results from
cool gas in SNRs (as explored in this paper), then essentially all AGNs
must have a {\em nuclear} starburst.

We have assumed in our models that both the ejecta, and the surrounding interstellar 
medium, are homogeneous and have solar abundances. 
While the presence of large-scale macroscopic mixing of ejecta in core collapse
SNe has been well established on both observational and theoretical grounds
(see Blondin \etal \cite{BBR2001} and references therein), such
mixing is not complete on a microscopic level. For example, X-ray observations 
of the Vela SNR (Aschenbach \etal \cite{AET1995}) show several fragments 
outside of the general boundary, and {\it ASCA} (Tsunemi \etal \cite{TMA1999})
and {\it XMM-Newton} (Aschenbach \& Miyata \cite{AM2003}) observations of
fragment A have revealed a significant overabundance of Si and Mg, confirming
that this fragment is ejecta. Widespread evidence that the ejecta of
core collapse supernovae are clumpy is further noted in Wang \& Chevalier (\cite{WC2002}).
The possibility that supernovae are explosions of ``shrapnel'' which give rise
to a complex outer boundary has been discussed by Kundt (\cite{K1988}).
Density enhancements in the surrounding medium may also occur. 
The interaction of ejecta with a clumpy wind has been
proposed as the origin of the broad and intermediate-width lines in the
spectrum of the peculiar SN~1988Z (Chugai \& Danziger \cite{CD1994}).

Knots of X-ray emission seen in the Tycho SNR (a type Ia explosion) also 
indicate clumpy ejecta, and abundance variations between the knots indicate
that the mixing of the deep Fe layer changes from place to place 
(Decourchelle \etal \cite{DSA2001}). However,
on a larger scale a general stratification is seen, with the lighter elements
found predominantly at large radii, and vice-versa. Emission from Fe has the
smallest radius, confirming the onion shell structure expected from 
deflagration models.

While this work has
used solar abundances and homogeneous media for simplicity, it is clear 
that the majority of the cool mass will be nuclear processed material
and that there will be localized density enhancements in the ejecta. 
Such material will cool more efficiently, and higher masses for the cold phase 
will be obtained. In this sense, the values in Table~\ref{tab:mass_cool_gas}
should be viewed as lower limits. Future models will eventually need to
treat in a realistic fashion the inhomogeneities in density and abundance that 
we know exist.

Our investigation of the influence of an AGN environment on the dynamics 
and evolution of a supernova remnant is ongoing.
In the next paper in this series we will study the {\em dynamical} influence of the 
QSO radiation field. We also note that the combined wind from a group 
of early-type stars may provide the necessary conditions for the
formation of cool regions. We anticipate that this scenario will be more 
relevant in the nuclei of Seyfert galaxies, since supernova explosions
will evacuate all but the most tightly bound gas in them (Perry \& Dyson
\cite{PD1985}). Finally, it is clear from our models that while the
supernova-QSO wind interaction is conceptually simple, the BELR
is likely to be a very complicated region in practice.

\begin{acknowledgements}
We would like to acknowledge helpful comments from the referee.
JMP would also like to thank PPARC for the funding of a PDRA position. 
Finally, we would like to give particular thanks to T. Woods for the use 
of his cooling and heating tables, and for the other help that he has kindly given
during the course of this work. This research has made use of NASA's Astrophysics 
Data System Abstract Service. 
\end{acknowledgements}

\end{document}